ARTICLE  OPEN

# Scalable and accurate deep learning with electronic health records

Alvin Rajkomar[1,2], Eyal Oren[1], Kai Chen[1], Andrew M. Dai[1], Nissan Hajaj[1], Michaela Hardt[1], Peter J. Liu[1], Xiaobing Liu[1], Jake Marcus[1], Mimi Sun[1], Patrik Sundberg[1], Hector Yee[1], Kun Zhang[1], Yi Zhang[1], Gerardo Flores[1], Gavin E. Duggan[1], Jamie Irvine[1], Quoc Le[1], Kurt Litsch[1], Alexander Mossin[1], Justin Tansuwan[1], De Wang[1], James Wexler[1], Jimbo Wilson[1], Dana Ludwig[2], Samuel L. Volchenboum[3], Katherine Chou[1], Michael Pearson[1], Srinivasan Madabushi[1], Nigam H. Shah[4], Atul J. Butte[2], Michael D. Howell[1], Claire Cui[1], Greg S. Corrado[1] and Jeffrey Dean[1]

Predictive modeling with electronic health record (EHR) data is anticipated to drive personalized medicine and improve healthcare quality. Constructing predictive statistical models typically requires extraction of curated predictor variables from normalized EHR data, a labor-intensive process that discards the vast majority of information in each patient's record. We propose a representation of patients' entire raw EHR records based on the Fast Healthcare Interoperability Resources (FHIR) format. We demonstrate that deep learning methods using this representation are capable of accurately predicting multiple medical events from multiple centers without site-specific data harmonization. We validated our approach using de-identified EHR data from two US academic medical centers with 216,221 adult patients hospitalized for at least 24 h. In the sequential format we propose, this volume of EHR data unrolled into a total of 46,864,534,945 data points, including clinical notes. Deep learning models achieved high accuracy for tasks such as predicting: in-hospital mortality (area under the receiver operator curve [AUROC] across sites 0.93–0.94), 30-day unplanned readmission (AUROC 0.75–0.76), prolonged length of stay (AUROC 0.85–0.86), and all of a patient's final discharge diagnoses (frequency-weighted AUROC 0.90). These models outperformed traditional, clinically-used predictive models in all cases. We believe that this approach can be used to create accurate and scalable predictions for a variety of clinical scenarios. In a case study of a particular prediction, we demonstrate that neural networks can be used to identify relevant information from the patient's chart.

npj Digital Medicine (2018)1:18 ; doi:10.1038/s41746-018-0029-1

## INTRODUCTION

The promise of digital medicine stems in part from the hope that, by digitizing health data, we might more easily leverage computer information systems to understand and improve care. In fact, routinely collected patient healthcare data are now approaching the genomic scale in volume and complexity.[1] Unfortunately, most of this information is not yet used in the sorts of predictive statistical models clinicians might use to improve care delivery. It is widely suspected that use of such efforts, if successful, could provide major benefits not only for patient safety and quality but also in reducing healthcare costs.[2–6]

In spite of the richness and potential of available data, scaling the development of predictive models is difficult because, for traditional predictive modeling techniques, each outcome to be predicted requires the creation of a custom dataset with specific variables.[7] It is widely held that 80% of the effort in an analytic model is preprocessing, merging, customizing, and cleaning datasets,[8,9] not analyzing them for insights. This profoundly limits the scalability of predictive models.

Another challenge is that the number of potential predictor variables in the electronic health record (EHR) may easily number in the thousands, particularly if free-text notes from doctors, nurses, and other providers are included. Traditional modeling approaches have dealt with this complexity simply by choosing a very limited number of commonly collected variables to consider.[7] This is problematic because the resulting models may produce imprecise predictions: false-positive predictions can overwhelm physicians, nurses, and other providers with false alarms and concomitant alert fatigue,[10] which the Joint Commission identified as a national patient safety priority in 2014.[11] False-negative predictions can miss significant numbers of clinically important events, leading to poor clinical outcomes.[11,12] Incorporating the entire EHR, including clinicians' free-text notes, offers some hope of overcoming these shortcomings but is unwieldy for most predictive modeling techniques.

Recent developments in deep learning and artificial neural networks may allow us to address many of these challenges and unlock the information in the EHR. Deep learning emerged as the preferred machine learning approach in machine perception problems ranging from computer vision to speech recognition, but has more recently proven useful in natural language processing, sequence prediction, and mixed modality data settings.[13–17] These systems are known for their ability to handle large volumes of relatively messy data, including errors in labels and large numbers of input variables. A key advantage is that investigators do not generally need to specify which potential predictor variables to consider and in what combinations; instead neural networks are able to learn representations of the key factors and interactions from the data itself.

We hypothesized that these techniques would translate well to healthcare; specifically, deep learning approaches could incorporate the entire EHR, including free-text notes, to produce predictions for a wide range of clinical problems and outcomes that outperform state-of-the-art traditional predictive models. Our central insight was that rather than explicitly harmonizing EHR data, mapping it into a highly curated set of structured predictors variables and then feeding those variables into a statistical model, we could instead learn to simultaneously harmonize inputs and predict medical events through direct feature learning.[18]

Related work

The idea of using computer systems to learn from a "highly organized and recorded database" of clinical data has a long history.[19] Despite the rich data now digitized in EHRs,[20] a recent systematic review of the medical literature[7] found that predictive models built with EHR data use a median of only 27 variables, rely on traditional generalized linear models, and are built using data at a single center. In clinical practice, simpler models are most commonly deployed, such as the CURB-65,[21,22] which is a 5-factor model, or single-parameter warning scores.[23,24]

A major challenge in using more of the data available for each patient has been the lack of standards and semantic interoperability of health data from multiple sites.[25] A unique set of variables is typically selected for each new prediction task, and usually a labor-intensive[8,9] process is required to extract and normalize data from different sites.[26]

Significant prior research has focused on the scalability issue through time-consuming standardization of data in traditional relational databases, like the Observational Medical Outcomes Partnership standard defined by the Observational Health Data Sciences and Informatics consortium.[27] Such a standard allows for consistent development of predictive models across sites, but accommodates only a part of the original data.

Recently, a flexible data structure called FHIR (Fast Healthcare Interoperability Resources)[28] was developed to represent clinical data in a consistent, hierarchical, and extensible container format, regardless of the health system, which simplifies data interchange between sites. However, the format does not ensure semantic consistency, motivating the need for additional techniques to deal with unharmonized data.

The use of deep learning on EHR data burgeoned after adoption of EHRs[20] and development of deep learning methods.[13] In a well-known work, investigators used auto-encoders to predict a specific set of diagnoses.[29] Subsequent work extended this approach by modeling the temporal sequence of events that occurred in a patient's record, which may enhance accuracy in scenarios that depend on the order of events, with convolutional and recurrent neural networks.[30–35] In general, prior work has focused on a subset of features available in the EHR, rather than on all data available in an EHR, which includes clinical free-text notes, as well as large amounts of structured and semi-structured data. Because of the availability of Medical Information Mart for Intensive Care (MIMIC) data,[36] many prior studies also have focused on ICU patients from a single center;[33,37] other single-center studies have also focused on ICU patients.[30] Each ICU patient has significantly more data available than each general hospital patient, although non-ICU admissions outnumber ICU admissions by about sixfold in the US.[38,39] Recently, investigators have also explored how interpretation mechanisms for deep learning models could be applied to clinical predictions.[33] Given rapid developments in this field, we point readers to a recent, comprehensive review.[40]

Our contribution is twofold. First, we report a generic data processing pipeline that can take raw EHR data as input, and produce FHIR outputs without manual feature harmonization. This makes it relatively easy to deploy our system to a new hospital. Second, based on data from two academic hospitals with a general patient population (not restricted to ICU), we demonstrate the effectiveness of deep learning models in a wide variety of predictive problems and settings (e.g., multiple prediction timing). Ours is a comprehensive study of deep learning in a variety of prediction problems based on multiple general hospital data. We do note, however, that similar deep learning techniques have been applied to EHR data in prior research as described above.

RESULTS

We included a total of 216,221 hospitalizations involving 114,003 unique patients. The percent of hospitalizations with in-hospital deaths was 2.3% (4930/216,221), unplanned 30-day readmissions was 12.9% (27,918/216,221), and long length of stay was (23.9%). Patients had a range of 1–228 discharge diagnoses. The demographics and utilization characteristics are summarized in Table 1. The median duration of patients' records, calculated by the difference of the timestamps of last and first FHIR resource was 3.1 years in Hospital A and 3.6 years in Hospital B.

At the time of admission, an average admission had 137,882 tokens (discrete pieces of data that we define in the methods section), which increased markedly throughout the patient's stay to 216,744 at discharge (Fig. 1). For predictions made at discharge, the information considered across both datasets included 46,864,534,945 tokens of EHR data.

Mortality

For predicting inpatient mortality, the area under the receiver operating characteristic curve (AUROC) at 24 h after admission was 0.95 (95% CI 0.94–0.96) for Hospital A and 0.93 (95% CI 0.92–0.94) for Hospital B. This was significantly more accurate than the traditional predictive model, the augmented Early Warning Score (aEWS) which was a 28-factor logistic regression model (AUROC 0.85 (95% CI 0.81–0.89) for Hospital A and 0.86 (95% CI 0.83–0.88) for Hospital B) (Table 2).

If a clinical team had to investigate patients predicted to be at high risk of dying, the rate of false alerts at each point in time was roughly halved by our model: at 24 h, the work-up-to-detection ratio of our model compared to the aEWS was 7.4 vs 14.3 (Hospital A) and 8.0 vs 15.4 (Hospital B). Moreover, the deep learning model achieved higher discrimination at every prediction time-point compared to the baseline models. The deep learning model attained a similar level of accuracy at 24–48 h earlier than the traditional models (Fig. 2).

Readmissions

For predicting unexpected readmissions within 30 days, the AUROCs at discharge were 0.77 (95% CI 0.75–0.78) for Hospital A and 0.76 (95% CI 0.75–0.77) for Hospital B. These were significantly higher than the traditional predictive model (modified HOSPITAL) at discharge, which were 0.70 (95% CI 0.68–0.72) for Hospital A and 0.68 (95% CI 0.67–0.69) for Hospital B.

Long length of stay

For predicting long length of stay, the AUROCs at 24 h after admission were 0.86 (95% CI 0.86–0.87) for Hospital A and 0.85 (95% CI 0.84–0.86) for Hospital B. These were significantly higher than those from the traditional predictive model (modified Liu) at

| Table 1. Characteristics of hospitalizations in training and test sets | | | | |
|---|---|---|---|---|
| | Training data (n = 194,470) | | Test data (n = 21,751) | |
| | Hospital A (n = 85,522) | Hospital B (n = 108,948) | Hospital A (n = 9624) | Hospital B (n = 12,127) |
| *Demographics* | | | | |
| Age, median (IQR) y | 56 (29) | 57 (29) | 55 (29) | 57 (30) |
| Female sex, no. (%) | 46,848 (54.8%) | 62,004 (56.9%) | 5364 (55.7%) | 6935 (57.2%) |
| *Disease cohort, no. (%)* | | | | |
| Medical | 46,579 (54.5%) | 55,087 (50.6%) | 5263 (54.7%) | 6112 (50.4%) |
| Cardiovascular | 4616 (5.4%) | 6903 (6.3%) | 528 (5.5%) | 749 (6.2%) |
| Cardiopulmonary | 3498 (4.1%) | 9028 (8.3%) | 388 (4.0%) | 1102 (9.1%) |
| Neurology | 6247 (7.3%) | 6653 (6.1%) | 697 (7.2%) | 736 (6.1%) |
| Cancer | 14,544 (17.0%) | 19,328 (17.7%) | 1617 (16.8%) | 2087 (17.2%) |
| Psychiatry | 788 (0.9%) | 339 (0.3%) | 64 (0.7%) | 35 (0.3%) |
| Obstetrics and newborn | 8997 (10.5%) | 10,462 (9.6%) | 1036 (10.8%) | 1184 (9.8%) |
| Other | 253 (0.3%) | 1148 (1.1%) | 31 (0.3%) | 122 (1.0%) |
| *Previous hospitalizations, no. (%)* | | | | |
| 0 hospitalizations | 54,954 (64.3%) | 56,197 (51.6%) | 6123 (63.6%) | 6194 (51.1%) |
| ≥1 and <2 hospitalizations | 14,522 (17.0%) | 19,807 (18.2%) | 1620 (16.8%) | 2175 (17.9%) |
| ≥2 and <6 hospitalizations | 12,591 (14.7%) | 24,009 (22.0%) | 1412 (14.7%) | 2638 (21.8%) |
| ≥6 hospitalizations | 3455 (4.0%) | 8935 (8.2%) | 469 (4.9%) | 1120 (9.2%) |
| *Discharge location no. (%)* | | | | |
| Home | 70,040 (81.9%) | 91,273 (83.8%) | 7938 (82.5%) | 10,109 (83.4%) |
| Skilled nursing facility | 6601 (7.7%) | 5594 (5.1%) | 720 (7.5%) | 622 (5.1%) |
| Rehabilitation | 2666 (3.1%) | 5136 (4.7%) | 312 (3.2%) | 649 (5.4%) |
| Another healthcare facility | 2189 (2.6%) | 2052 (1.9%) | 243 (2.5%) | 220 (1.8%) |
| Expired | 1816 (2.1%) | 2679 (2.5%) | 170 (1.8%) | 265 (2.2%) |
| Other | 2210 (2.6%) | 2214 (2.0%) | 241 (2.5%) | 262 (2.2%) |
| *Primary outcomes* | | | | |
| In-hospital deaths, no. (%) | 1816 (2.1%) | 2679 (2.5%) | 170 (1.8%) | 265 (2.2%) |
| 30-day readmissions, no. (%) | 9136 (10.7%) | 15,932 (14.6%) | 1013 (10.5%) | 1837 (15.1%) |
| Hospital stays at least 7 days, no. (%) | 20,411 (23.9%) | 26,109 (24.0%) | 2145 (22.3%) | 2931 (24.2%) |
| No. of ICD-9 diagnoses, median (IQR) | 12 (16) | 10 (10) | 12 (16) | 10 (10) |

24 h, which were 0.76 (95% CI 0.75–0.77) for Hospital A and 0.74 (95% CI 0.73–0.75) for Hospital B.

Calibration curves for the three tasks are shown in Supplement.

Inferring discharge diagnoses

The deep learning algorithm predicted patients' discharge diagnoses at three time points: at admission, after 24 h of hospitalization, and at the time of discharge (but before the discharge diagnoses were coded). For classifying all diagnosis codes, the weighted AUROCs at admission were 0.87 for Hospital A and 0.86 for Hospital B. Accuracy increased somewhat during the hospitalization, to 0.88–0.89 at 24 h and 0.90 for both hospitals at discharge. For classifying ICD-9 code predictions as correct, we required full-length code agreement. For example, 250.4 ("Diabetes with renal manifestations") would be considered different from 250.42 ("Diabetes with renal manifestations, type II or unspecified type, uncontrolled"). We also calculated the micro-F1 scores at discharge, which were 0.41 (Hospital A) and 0.40 (Hospital B).

Case study of model interpretation

In Fig. 3, we illustrate an example of attribution methods on a specific prediction of inpatient mortality made at 24 h after admission. For this patient, the deep learning model predicted the risk of death of 19.9% and the aEWS model predicted 9.3%, and the patient ultimately died 10 days after admission. This patient's record had 175,639 data points (tokens), which were considered by the model. The timeline in Fig. 3 highlights the elements to which the model attends, with a close-up view of the first 24 h of the most recent hospitalization. From all the data, the models picked the elements that are highlighted in Fig. 3: evidence of malignant pleural effusions and empyema from notes, antibiotics administered, and nursing documentation of a high risk of pressure ulcers (i.e., Braden index[41]). The model also placed high weights on concepts, such as "pleurx," the trade name for a small chest tube. The bolded sections are exactly what the model identified as discriminatory factors, not a manual selection. In contrast, the top predictors for the baseline model (not shown in Fig. 3) were the values of the albumin, blood-urea-nitrogen, pulse, and white blood cell count. Note that for demonstration purposes, this example was generated from time-aware neural network models (TANNs) trained on separate modalities (e.g., flowsheets and notes), which is a common visualization technique to handle redundant features in the data (e.g., medication orders are also referenced in notes).

DISCUSSION

A deep learning approach that incorporated the entire EHR, including free-text notes, produced predictions for a wide range of clinical problems and outcomes and outperformed traditional,

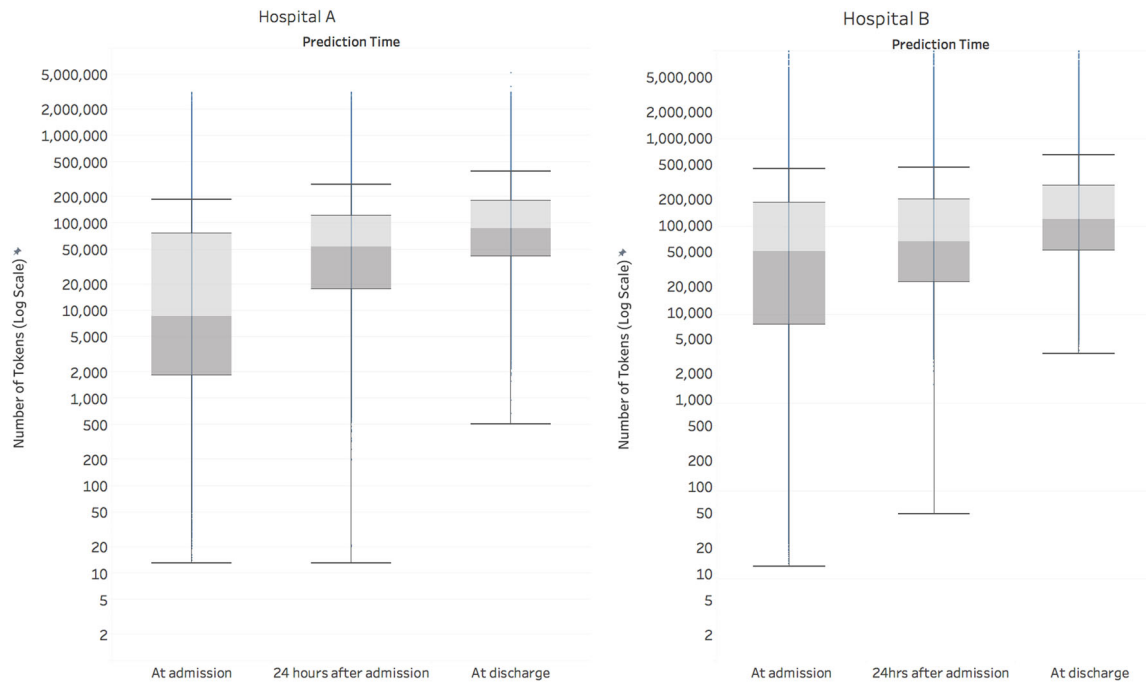

**Fig. 1** This boxplot displays the amount of data (on a log scale) in the EHR, along with its temporal variation across the course of an admission. We define a token as a single data element in the electronic health record, like a medication name, at a specific point in time. Each token is considered as a potential predictor by the deep learning model. The line within the boxplot represents the median, the box represents the interquartile range (IQR), and the whiskers are 1.5 times the IQR. The number of tokens increased steadily from admission to discharge. At discharge, the median number of tokens for Hospital A was 86,477 and for Hospital B was 122,961

| Table 2. Prediction accuracy of each task made at different time points | | |
|---|---|---|
| | Hospital A | Hospital B |
| *Inpatient mortality, AUROC[a] (95% CI)* | | |
| 24 h before admission | 0.87 (0.85–0.89) | 0.81 (0.79–0.83) |
| At admission | 0.90 (0.88–0.92) | 0.90 (0.86–0.91) |
| 24 h after admission | **0.95 (0.94–0.96)** | **0.93 (0.92–0.94)** |
| Baseline (aEWS[b]) at 24 h after admission | 0.85 (0.81–0.89) | 0.86 (0.83–0.88) |
| *30-day readmission, AUROC (95% CI)* | | |
| At admission | 0.73 (0.71–0.74) | 0.72 (0.71–0.73) |
| At 24 h after admission | 0.74 (0.72–0.75) | 0.73 (0.72–0.74) |
| At discharge | **0.77 (0.75–0.78)** | **0.76 (0.75–0.77)** |
| Baseline (mHOSPITAL[c]) at discharge | 0.70 (0.68–0.72) | 0.68 (0.67–0.69) |
| *Length of stay at least 7 days, AUROC (95% CI)* | | |
| At admission | 0.81 (0.80–0.82) | 0.80 (0.80–0.81) |
| At 24 h after admission | **0.86 (0.86–0.87)** | **0.85 (0.85–0.86)** |
| Baseline (Liu[d]) at 24 h after admission | 0.76 (0.75–0.77) | 0.74 (0.73–0.75) |
| *Discharge diagnoses (weighted AUROC)* | | |
| At admission | 0.87 | 0.86 |
| At 24 h after admission | 0.89 | 0.88 |
| At discharge | **0.90** | **0.90** |

[a]Area under the receiver operator curve
[b]Augmented Early Warning System score
[c]Modified HOSPITAL score for readmission
[d]Modified Liu score for long length of stay
The bold values indicate the highest area-under-the-receiver-operator-curve for each prediction task clinically-used predictive models. Because we were interested in understanding whether deep learning could scale to produce valid predictions across divergent healthcare domains, we used a single data structure to make predictions for an important clinical outcome (death), a standard measure of quality of care (readmissions), a measure of resource utilization (length of stay), and a measure of understanding of a patient's problems (diagnoses).

Second, using the entirety of a patient's chart for every prediction does more than promote scalability, it exposes more data with which to make an accurate prediction. For predictions made at discharge, our deep learning models considered more than 46 billion pieces of EHR data and achieved more accurate predictions, earlier in the hospital stay, than did traditional models.

To the best of our knowledge, our models outperform existing EHR models in the medical literature for predicting mortality (0.92–0.94 vs 0.91),[42] unexpected readmission (0.75–0.76 vs 0.69),[43] and increased length of stay (0.85–0.86 vs 0.77).[44] Direct comparisons to other studies are difficult[45] because of different underlying study designs,[23,46–57] incomplete definitions of cohorts and outcomes,[58,59] restrictions on disease-specific cohorts[58–64], or use of data unavailable in real-time.[63,65,66] Therefore, we implemented baselines based on the HOSPITAL score,[67] NEWS[51] score, and Liu's model[44] on our data, and demonstrate strictly better performance. We are not aware of a study that predicts as many ICD codes as this study, but our micro-F1 score exceeds that shown on the smaller MIMIC-III dataset when predicting fewer diagnoses (0.40 vs 0.28).[68] The clinical impact of this improvement is suggested, for example, by the improvement of number needed to evaluate for inpatient mortality: the deep learning model would fire half the number of alerts of a traditional predictive model, resulting in many fewer false positives.

However, the novelty of the approach does not lie simply in incremental model performance improvements. Rather, this predictive performance was achieved without hand-selection of variables deemed important by an expert, similar to other

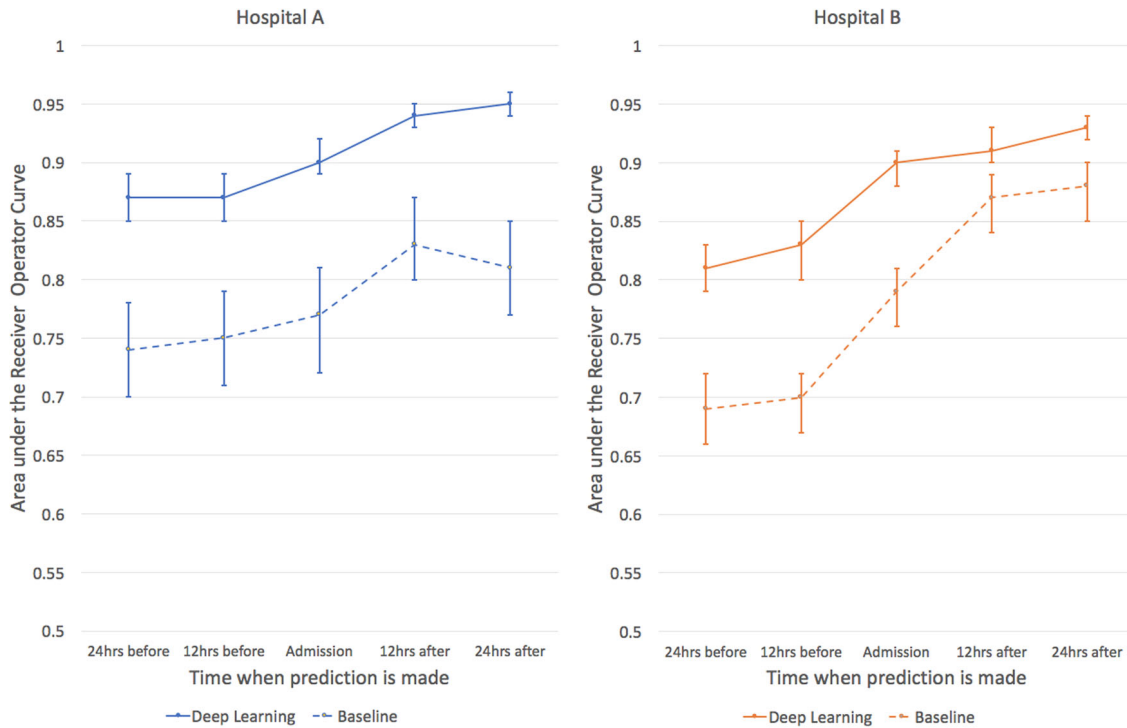

**Fig. 2** The area under the receiver operating characteristic curves are shown for predictions of inpatient mortality made by deep learning and baseline models at 12 h increments before and after hospital admission. For inpatient mortality, the deep learning model achieves higher discrimination at every prediction time compared to the baseline for both the University of California, San Francisco (UCSF) and University of Chicago Medicine (UCM) cohorts. Both models improve in the first 24 h, but the deep learning model achieves a similar level of accuracy approximately 24 h earlier for UCM and even 48 h earlier for UCSF. The error bars represent the bootstrapped 95% confidence interval

applications of deep learning to EHR data. Instead, our model had access to tens of thousands of predictors for each patient, including free-text notes, and identified which data were important for a particular prediction.

Our study also has important limitations. First, it is a retrospective study, with all of the usual limitations. Second, although it is widely believed that accurate predictions can be used to improve care,[4] this is not a foregone conclusion and prospective trials are needed to demonstrate this.[69,70] Third, a necessary implication of personalized predictions is that they leverage many small data points specific to a particular EHR rather than a handful of common variables. Future research is needed to determine how models trained at one site can be best applied to another site,[71] which would be especially useful for sites with limited historical data for training. As a first step, we demonstrated that similar model architectures and training methods yielded comparable models for two geographically distinct health systems. Our current approach does not harmonize data between sites, which limits the model's ability to "transfer" learn from one site to other sites and cohorts, and further research is needed. Moreover, our methods are computationally intensive and at present require specialized expertize to develop, but running the models on a new patient takes only a few milliseconds. The availability and accessibility of machine learning is also rapidly expanding both in healthcare and in other fields. Another limitation is that the current study focuses on predictive accuracy as a whole rather than incremental benefit of a given data type (e.g., clinical notes). We view understanding the incremental contribution of notes to predictive performance as an important area of future investigation, including identifying tasks and metrics where notes should have significant impact, testing different approaches to modeling the note terms, and understanding whether different portions of a note have different contributions to predictive accuracy. In the current study, we caution that the differences in AUROC across the two hospitals

(one with and one without notes) cannot be ascribed to the presence or absence of notes given the difference in cohorts.[45]

Perhaps the most challenging prediction in our study is that of predicting a patient's full suite of discharge diagnoses. The prediction is difficult for several reasons. First, a patient may have between 1 and 228 diagnoses, and the number is not known at the time of prediction. Second, each diagnosis may be selected from approximately 14,025 ICD-9 diagnosis codes, which makes the total number of possible combinations exponentially large. Finally, many ICD-9 codes are clinically similar but numerically distinct (e.g., 011.30 "Tuberculosis of bronchus, unspecified" vs 011.31 "Tuberculosis of bronchus, bacteriological or histological examination not done"). This has the effect of introducing random error into the prediction. The micro-F1 score, which is a metric used when a prediction has more than a single outcome (e.g., multiple diagnoses), for our model is higher than that reported in the literature in an ICU dataset with fewer diagnoses.[68] This is a proof-of-concept that demonstrates that the diagnosis could be inferred from routine EHR data, which could aid with triggering of decision support[68,72] or clinical trial recruitment.

The use of free text for prediction allows a new level of explainability of predictions. Clinicians have historically distrusted neural network models because of their opaqueness. We demonstrate how our method can visualize what data the model "looked at" for each individual patient, which can be used by a clinician to determine if a prediction was based on credible facts, and potentially help decide actions. In our case study, the model identified elements of the patient's history and radiology findings to render its prediction, which are critical data points that a clinician would also use.[73] This approach may address concerns that such "black box" methods are untrustworthy. However, there are other possible techniques for interpreting deep learning models.[33,74] We report the case study as a proof-of-concept drawn directly from our model architecture and data and emphasize that further research is needed regarding applicability to all predictions, the cognitive impact, and clinical utility.

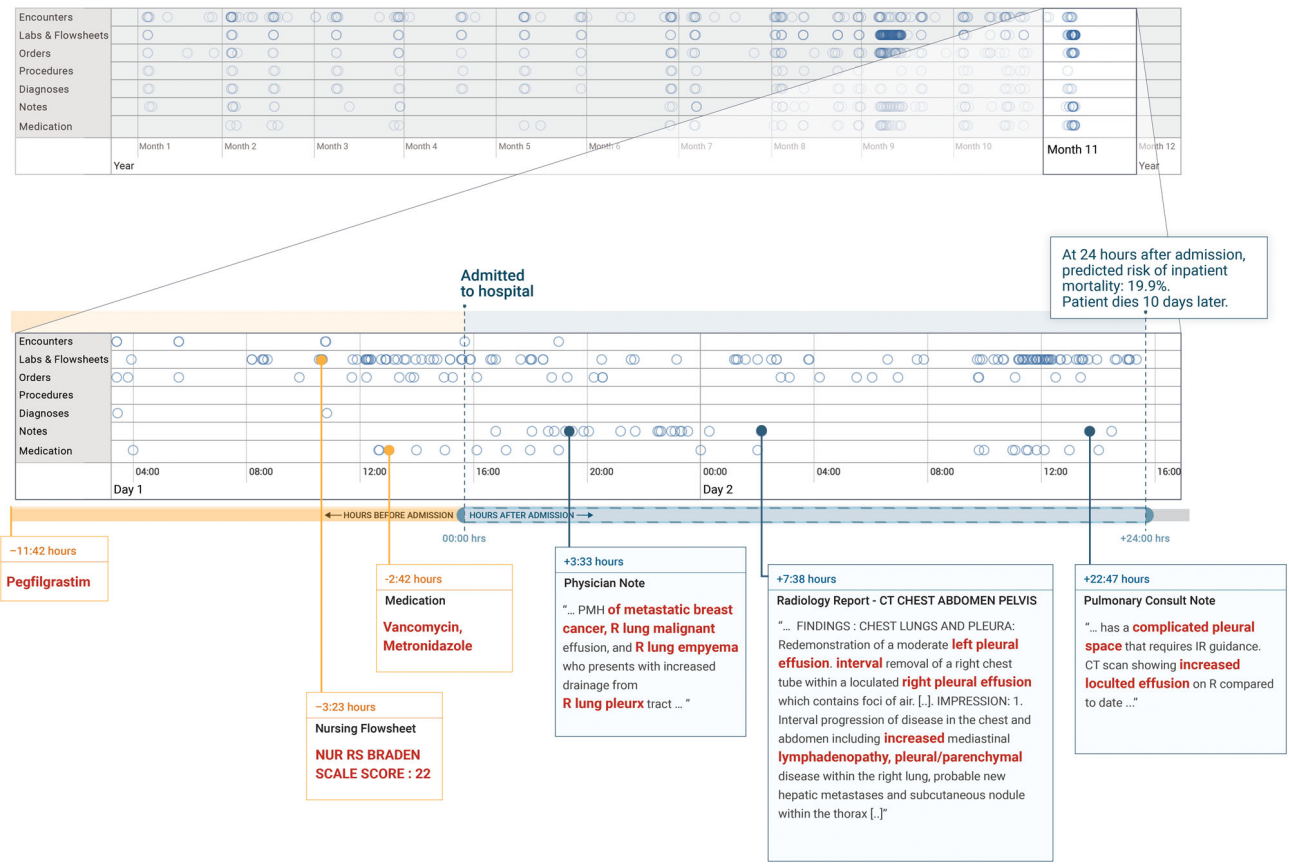

**Fig. 3** The patient record shows a woman with metastatic breast cancer with malignant pleural effusions and empyema. The patient timeline at the top of the figure contains circles for every time-step for which at least a single token exists for the patient, and the horizontal lines show the data type. There is a close-up view of the most recent data points immediately preceding a prediction made 24 h after admission. We trained models for each data type and highlighted in red the tokens which the models attended to—the non-highlighted text was not attended to but is shown for context. The models pick up features in the medications, nursing flowsheets, and clinical notes relevant to the prediction

## METHODS

### Datasets
We included EHR data from the University of California, San Francisco (UCSF) from 2012 to 2016, and the University of Chicago Medicine (UCM) from 2009 to 2016. We refer to each health system as Hospital A and Hospital B. All EHRs were de-identified, except that dates of service were maintained in the UCM dataset. Both datasets contained patient demographics, provider orders, diagnoses, procedures, medications, laboratory values, vital signs, and flowsheet data, which represent all other structured data elements (e.g., nursing flowsheets), from all inpatient and outpatient encounters. The UCM dataset additionally contained de-identified, free-text medical notes. Each dataset was kept in an encrypted, access-controlled, and audited sandbox.

Ethics review and institutional review boards approved the study with waiver of informed consent or exemption at each institution.

### Data representation and processing
We developed a single data structure that could be used for all predictions, rather than requiring custom, hand-created datasets for every new prediction. This approach represents the entire EHR in temporal order: data are organized by patient and by time. To represent events in a patient's timeline, we adopted the FHIR standard.[75] FHIR defines the high-level representation of healthcare data in resources, but leaves values in each individual site's idiosyncratic codings.[28] Each event is derived from a FHIR resource and may contain multiple attributes; for example, a medication-order resource could contain the trade name, generic name, ingredients, and others. Data in each attribute were split into discrete values, which we refer to as tokens. For notes, the text was split into a sequence of tokens, one for each word. Numeric values were normalized, as detailed in the supplement. The entire sequence of time-ordered tokens, from the beginning of a patient's record until the point of prediction, formed the patient's personalized input to the model. This process is illustrated in Fig. 4, and further details of the FHIR representation and processing are provided in Supplementary Materials.

### Outcomes
We were interested in understanding whether deep learning could produce valid predictions across wide range of clinical problems and outcomes. We therefore selected outcomes from divergent domains, including an important clinical outcome (death), a standard measure of quality of care (readmissions), a measure of resource utilization (length of stay), and a measure of understanding of a patient's problems (diagnoses).

*Inpatient mortality.* We predicted impending inpatient death, defined as a discharge disposition of "expired."[42,46,48,49]

*30-day unplanned readmission.* We predicted unplanned 30-day readmission, defined as an admission within 30 days after discharge from an "index" hospitalization. A hospitalization was considered a "readmission" if its admission date was within 30 days after discharge of an eligible index hospitalization. A readmission could only be counted once. There is no standard definition of "unplanned"[76] percentage, so we used a modified form of the Centers for Medicare and Medicaid Services definition,[77] which we detail in the supplement. Billing diagnoses and procedures from the index hospitalization were not used for the prediction because they are typically generated after discharge. We included only readmissions to the same institution.

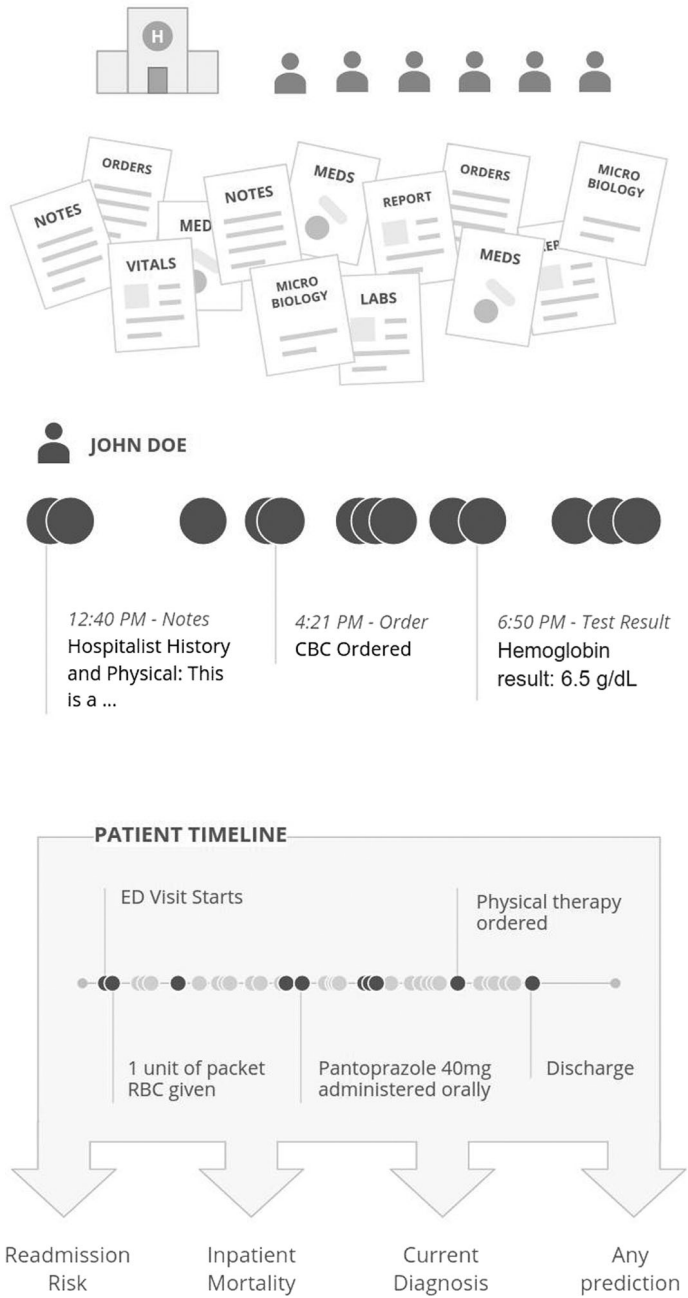

**Fig. 4** Data from each health system were mapped to an appropriate FHIR (Fast Healthcare Interoperability Resources) resource and placed in temporal order. This conversion did not harmonize or standardize the data from each health system other than map them to the appropriate resource. The deep learning model could use all data available prior to the point when the prediction was made. Therefore, each prediction, regardless of the task, used the same data

*Long length of stay.* We predicted a length of stay at least 7 days, which was approximately the 75th percentile of hospital stays for most services across the datasets. The length of stay was defined as the time between hospital admission and discharge.

*Diagnoses.* We predicted the entire set of primary and secondary ICD-9 billing diagnoses from a universe of 14,025 codes.

### Prediction timing
This was a retrospective study. To predict inpatient mortality, we stepped forward through each patient's time course, and made predictions every 12 h starting 24 h before admission until 24 h after admission. Since many clinical prediction models, such as APACHE,[78] are rendered 24 h after admission, our primary outcome prediction for inpatient mortality was at that time-point. Unplanned readmission and the set of diagnosis codes were predicted at admission, 24 h after admission, and at discharge. The primary endpoints for those predictions were at discharge, when most readmission prediction scores are computed[79] and when all information necessary to assign billing diagnoses is available. Long length of stay was predicted at admission and 24 h after admission. For every prediction we used all information available in the EHR up to the time at which the prediction was made.

### Study cohort

We included all admissions for patients 18 years or older. We only included hospitalizations of 24 h or longer to ensure that predictions at various time points had identical cohorts.

To simulate the accuracy of a real-time prediction system, we included patients typically removed in studies of readmission, such as those discharged against medical advice, since these exclusion criteria would not be known when making predictions earlier in the hospitalization.

For predicting the ICD-9 diagnoses, we excluded encounters without any ICD-9 diagnosis (2–12% of encounters). These were generally encounters after October, 2015 when hospitals switched to ICD-10. We included such hospitalizations, however, for all other predictions.

### Algorithm development and analysis

We used the same modeling algorithm on both hospitals' datasets, but treated each hospital as a separate dataset and reported results separately.

Patient records vary significantly in length and density of data points (e.g., vital sign measurements in an intensive care unit vs outpatient clinic), so we formulated three deep learning neural network model architectures that take advantage of such data in different ways: one based on recurrent neural networks (long short-term memory (LSTM)),[80] one on an attention-based TANN, and one on a neural network with boosted time-based decision stumps. Details of these architectures are explained in the supplement. We trained each architecture (three different ones) on each task (four tasks) and multiple time points (e.g., before admission, at admission, 24 h after admission and at discharge), but the results of each architecture were combined using ensembling.[81]

### Comparison to previously published algorithms

We implemented models based on previously published algorithms to establish baseline performance on each dataset. For mortality, we used a logistic model with variables inspired by NEWS[51] score but added additional variables to make it more accurate, including the most recent systolic blood pressure, heart rate, respiratory rate, temperature, and 24 common lab tests, like the white blood cell count, lactate, and creatinine. We call this the augmented Early Warning Score, or aEWS, score. For readmission, we used a logistic model with variables used by the HOSPITAL[67] score, including the most recent sodium and hemoglobin level, hospital service, occurrence of CPT codes, number of prior hospitalizations, and length of the current hospitalization. We refer to this as the mHOSPITAL score. For long length of stay, we used a logistic model with variables similar to those used by Liu:[44] the age, gender, hierarchical condition categories, admission source, hospital service, and the same 24 common lab tests used in the aEWS score. We refer to this as the mLiu score. Details for these and additional baseline models are in the supplement. We are not aware of any commonly used baseline model for all diagnosis codes so we compare against known literature.

### Explanation of predictions

A common criticism of neural networks is that they offer little insight into the factors that influence the prediction.[82] Therefore, we used attribution mechanisms to highlight, for each patient, the data elements that influenced their predictions.[83]

The LSTM and TANN models were trained with TensorFlow and the boosting model was implemented with C++ code. Statistical analyses and baseline models were done in Scikit-learn Python.[84]

Technical details of the model architecture, training, variables, baseline models, and attribution methods are provided in the supplement.

### Model evaluation and statistical analysis

Patients were randomly split into development (80%), validation (10%), and test (10%) sets. Model accuracy is reported on the test set, and 1000 bootstrapped samples were used to calculate 95% confidence intervals. To prevent overfitting, the test set remained unused (and hidden) until final evaluation.

We assessed model discrimination by calculating AUROC and model calibration using comparisons of predicted and empirical probability curves.[85] We did not use the Hosmer–Lemeshow test as it may be misleadingly significant with large sample sizes.[86] To quantify the potential clinical impact of an alert with 80% sensitivity, we report the work-up to detection ratio, also known as the number needed to evaluate.[87] For prediction of the a patient's full set of diagnosis codes, which can range from 1 to 228 codes per hospitalization, we evaluated the accuracy for each class using macro-weighted-AUROC[88] and micro-weighted F1 score[89] to compare with the literature. The F1 score is the harmonic mean of positive-predictive-value and sensitivity; we used a single threshold picked on the validation set for all classes. We did not create confidence intervals for this task given the computational complexity of the number of possible diagnoses.

### Data availability

The datasets analysed during the current study are not publicly available: due to reasonable privacy and security concerns, the underlying EHR data are not easily redistributable to researchers other than those engaged in the Institutional Review Board-approved research collaborations with the named medical centers.

### Code availability

The FHIR format used in this work is available at https://github.com/google/fhir. The transformation of FHIR-formatted data to Tensorflow training examples and the models themselves depend on Google's internal distributed computation platforms that cannot be reasonably shared. We have therefore emphasized detailed description of how our models were constructed and designed in our Methods section and Supplementary Materials.

### ACKNOWLEDGEMENTS

For data acquisition and validation, we thank the following: Julie Johnson, Sharon Markman, Thomas Sutton, Brian Furner, Timothy Holper, Sharat Israni, Jeff Love, Doris Wong, Michael Halaas, and Juan Banda. For statistical consultation, we thank Farzan Rohani. For modeling infrastructure, we thank Daniel Hurt and Zhifeng Chen. For help with visualizing Figs. 1 and 3, we thank Kaye Mao and Mahima Pushkarna. Each organization supported its work using internal funding.

### AUTHOR CONTRIBUTIONS

A.R., E.O., and C.C. contributed with study design, data-cleaning, statistical analysis, interpretation of results, and drafted and revised the paper. K.C., A.M.D., N.H., P.J.L., X. L., M.S., P.S., H.Y., K.Z., and Y.Z. (ordered alphabetically) contributed with data processing, statistical analysis, machine learning, and revised the paper draft. G.E.D., K.L., A.M., J.T., D.W., J.W., and J.W. contributed with data infrastructure and processing and revised the paper draft. G.F., M.H., J.M., and J.I. contributed with data analysis and machine learning, and revised the paper draft. D.L., S.L.V. contributed with data collection, data analysis, and revised the paper draft. N.H.S. and A.J.B. contributed with study design, interpretation of results, and revised the paper draft. M.H. contributed with study design, interpretation of results, and drafted and revised the paper draft. K.C., M.P., and S.M. contributed with acquisition of data, and revised the paper draft. G.C. and J.D. contributed with study design, statistical analysis, machine learning, and revised the paper draft.

### ADDITIONAL INFORMATION

**Supplementary information** accompanies the paper on the *npj Digital Medicine* website (https://doi.org/10.1038/s41746-018-0029-1).

**Competing interests:** The authors declare no competing interests.

**Publisher's note:** Springer Nature remains neutral with regard to jurisdictional claims in published maps and institutional affiliations.